\documentclass[12pt]{article}

\def\be{\begin{equation}}
\def\ee{\end{equation}}
\def\ba{\begin{eqnarray}}
\def\ea{\end{eqnarray}}
\def\la{\langle}
\def\ra{\rangle}
\def\a{\alpha}
\def\b{\beta}

\def\lo{\longrightarrow}

\usepackage{graphicx}
\begin{document}
\begin{titlepage}
\vspace{4cm}
\begin{center}{\Large \bf Thermal entanglement of spins in mean-field clusters}\\
\vspace{1cm}M. Asoudeh\footnote{email:asoudeh@mehr.sharif.edu},
\hspace{0.5cm} V. Karimipour \footnote{Corresponding author, email:vahid@sharif.edu}\\
\vspace{1cm} Department of Physics, Sharif University of Technology,\\
P.O. Box 11365-9161,\\ Tehran, Iran
\end{center}
\vskip 3cm
\begin{abstract}
We determine thermal entanglement in mean field clusters of $N$ spin
one-half particles interacting via the anisotropic Heisenberg
interaction, with and without external magnetic field.  For the
$xxx$ cluster in the absence of magnetic field we prove that only
the $N=2$ ferromagnetic cluster shows entanglement. An external
magnetic field $B$ can only entangle $xxx$ anti-ferromagnetic
clusters in certain regions of the $B-T$ plane. On the other hand,
the $xxz$ clusters of size $N>2$  are entangled only when the
interaction is ferromagnetic. Detailed dependence of the
entanglement on various parameters is investigated in each case.

\end{abstract}

\end{titlepage}

\vskip 3cm
\section{Introduction}\label{intro}
It is usually stated that at any finite non-zero temperature,
thermal fluctuations suppress quantum fluctuations and for that
reason the latter can exist only at absolute zero or very low
temperatures. With the progress in quantifying entanglement
\cite{woo} or quantum correlations, we have now a precise method for
determining exactly at what temperatures quantum correlations cease
to exist. This problem has been investigated extensively under the
name "Thermal Entanglement" in many body quantum systems \cite{niel,
abv, gbkv, ow, wz, bost, akHeis, akInhomo, kamta, canosa, sun, rig,
wangThreshold, wangxx, wangfusol}. The term is usually used to
specify the amount of entanglement or genuine quantum correlation
which exists between two parts of a system when the whole system is
in a state of thermal equilibrium. One is interested in how this
quantity depends on the coupling constants of the system and the
temperature. Usually there is a threshold temperature above which
entanglement vanishes. In some rare cases it is even observed that
raising the temperature first increases the quantum correlation and
then begins to diminish it \cite{abv}. In this way by a quantitative
treatment of the problem we
are able to sharpen our intuitive notions on the effect of temperature on quantum correlations.\\

In addition to these purely theoretical considerations, these works
are also motivated by the proposals in which spins in solid state
systems play the role of qubits \cite{kane, loss, divin1, burkard,
imamoglu, kavokin, wu} and the fact that thermal entanglement seems
to be stable against de-coherence and needs no controlled switching for its generation.\\

In this regard in the past few years many types of spin systems have
been studied, which include, the Heisenberg rings with ferromagnetic
and anti-ferromagnetic coupling in the absence and presence of
magnetic fields, small clusters of spins with different types of
isotropic and anisotropic coupling in homogeneous and in-homogeneous
magnetic fields. These are only a small sample of the works which
have been done in this direction. There are also other works which
take into account other forms of interactions between the
spins.\\

Obtaining exact results on thermal entanglement in spin systems on
arbitrary lattices is extremely difficult, since this generally
requires a knowledge of different correlation  functions which in
turn requires a knowledge of the full spectrum of the system.
However in certain systems with extra symmetries one can obtain some
partial and interesting results \cite{wz}.\\

In this paper we want to study in some considerable detail thermal
entanglement of spins in mean field clusters of arbitrary sizes.
These are clusters in which every spin is coupled to every other
spin with equal strength. Such an study has a two-fold motivation.
First, mean field systems are amenable to exact analytical treatment
since as we will see, one can determine the entanglement by the sole
knowledge of energy eigenvalues, i.e. the partition function, and
second they are the first approximation that one uses to study other
models. Therefore the results which are obtained for entanglement in
these systems, may be good approximations for their corresponding
results on other lattices
specially when the coordination number of the latter is high. \\

We would like to stress that in all these considerations we are
concerned with the thermal state of the system which at zero
temperature approaches an equal mixture of various ground states, if
there is a degeneracy in the spectrum. For the entanglement
properties of the ground states of
$xy$ mean field clusters the reader is referred to \cite{vidal1, vidal2, vidal3, vidal4, vidal5, vidal6}.\\
In these works a general class of mean field models under the name
collective models have been considered and the entanglement
properties of their ground states with particular emphasis to its
relation with quantum phase transitions have been studied.\\

The structure and the results of this paper are as follows: In
section \ref{model} we introduce the model which consists of a
system of $N$ spin one-half particles interacting with each other
and placed in an external magnetic field. The interaction may be of
ferromagnetic or anti-ferromagnetic type and there may also be
present some degree of anisotropy in the couplings of spins in
different directions. We obtain general formulas for the thermal
entanglement of this model derived from its partition function. In
the following sections we study in detail special cases of the
model, that is, in section \ref{xxx} we consider the isotropic (or
$xxx$ model) without magnetic field for both ferromagnetic and
anti-ferromagnetic interactions, where we show that only the
anti-ferromagnetic cluster of $N=2$ spins shows entanglement. In
section \ref{xxxB} we study the effect of magnetic field in the
$xxx$ model and show that the ferromagnetic clusters show no
entanglement while anti-ferromagnetic clusters have entanglement in
certain regions of the $B-T$ plane, the plane of magnetic field and
temperature. Finally in section \ref{xxz} we consider the
anisotropic ($xxz$ model) without magnetic field. For an $N=2$
cluster both the ferromagnetic and anti-ferromagnetic interactions
produce entanglement, however remarkably for $N>2$ we see that only
ferromagnetic clusters show entanglement and no entanglement
develops in anti-ferromagnetic clusters. We determine the regions in
the $\Delta-T$ plane, the plane of the anisotropy and temperature
where entanglement is non zero.  In all cases we determine the
dependence of the entanglement in terms of the size of the cluster.
Needless to say in a mean field cluster the entanglement between any
two spins is very low compared to the chain, due to the large number
of neighbors of a site. Therefore we use the re-scaled concurrence
\cite{vidal2}, the ordinary concurrence scaled by the number of
neighbors to measure the degree of entanglement of a cluster.

\section{Interacting spins on mean field graphs}\label{model}
Consider a cluster of spin $\frac{1}{2}$ particles in an external
magnetic field and interacting with each other. We take the
Hamiltonian as
\begin{equation}\label{H}
    H=\frac{J}{N-1}\sum_{i\ne j=1}^N (\vec{s}_{i}\cdot \vec{s}_j+\Delta
    s_{i z}s_{j z}) + B\sum_{i=1}^N s_{iz},
\end{equation}
where  $s_{i a}=\frac{1}{2}\sigma_{i a}$ and $\sigma_{i a}$'s are
Pauli operators. The coupling constant is taken as $\frac{J}{N-1}$
to ensure an extensive total energy. We also set $J=-1$ for
ferromagnetic
 and $J=+1$ for anti-ferromagnetic clusters and without loss of generality $B$ is taken positive.
 The parameter $\Delta$
 determines the anisotropy of the interaction. For $\Delta=0$ we
 have an isotropic ($xxx$ system which in the absence of external magnetic field
 displays full $SU(2)$ symmetry. This system is exactly solvable
 since one can rewrite it in terms of the total spin operators $S_a:=\sum_{i}{s_{i a}}$ in the form
 \begin{equation}\label{Htotal}
    H=\frac{J}{N-1}(\vec{S}\cdot \vec{S}+ \Delta
    S^2_z)+BS_z,
\end{equation}
where $S^2 = S_x^2+S_y^2+S_z^2$ and we have ignored an additive
constant. Note that in studying the ground state properties where
one can restrict to a fixed spin sector (i.e.
$S_x^2+S_y^2+S_z^2=S^2$), the model is equivalent to the
Lipkin-Meshkov-Glick model \cite{lmg} which has been extensively
studied in \cite{vidal3,vidal4,vidal5}. However for the thermal
entanglement properties of the model one should take into account
all spin sectors and the above equivalence no longer holds.

For any system of spins having the symmetry $[S_z, H]=0$, it has
been shown \cite{woo, wz} that the reduced density matrix between
two spins is given by
\begin{equation}\label{rho}
    \rho_{ij}=\left(\begin{array}{cccc} u^+& && \\ & w_1 & z^* & \\ & z & w_2 & \\ &&&
    u^-\end{array}\right),
\end{equation}
where
\begin{eqnarray}
  u^{\pm}&=& \frac{1}{4}\pm \mu +  G_{zz}, \\
  z &=& G_{xx}+G_{yy} +i(G_{xy}-G_{yx}),
\end{eqnarray}
and we have ignored the labels indicating the two sites in question
which for the mean field cluster can be any two sites.  Here
$G_{\a\b}$ and $\mu$ are the spin-spin correlation functions and the
magnetization per site respectively.
\begin{eqnarray}
  G_{\a\b} &=& \la s_{i\a}s_{j\b}\ra, \cr
  \mu &=& \la s_{iz}\ra =\frac{1}{N}\la S_{z}\ra.
\end{eqnarray}
(Note that there is slight difference between our notations and that
of \cite{wz} who take $G_{ab}=\la \sigma_a\sigma_b\ra$). For such a
density matrix the concurrence is given by \cite{woo, wz}
\begin{equation}\label{Cgeneral}
    C=2\ \ max \left[0, \mid z\mid - \sqrt{u^+u^-} \right].
\end{equation}
from which one can obtain the entanglement of formation (Eof) from
\begin{equation}\label{Eof}
    Eof(\rho)=-\lambda \log \lambda - (1-\lambda) \log (1-\lambda),
\end{equation}
with $\lambda = \frac{1-\sqrt{1-C^2}}{2}$. We now note that $z$
 is real. This can be proved by resorting to the reality of
 the Hamiltonian and hence the reality of the total and the reduced density
 matrices. It is worth mentioning that when the interaction strengths between the $x$, and the $y$ terms are
 not equal, the symmetry $[H,S_z]=0$ will no longer hold and the reduced density matrix will not be of the form
 (\ref{rho}). For these cases one can resort to a recent conjecture of \cite{vidal3} who provide the closed form of the concurrence for general
 models having permutation symmetry.  \\

Thus the concurrence can be expressed solely in terms of the
correlation functions
 \begin{equation}\label{C}
    C = 2\ \  max \left[0,  \mid  G_{xx}+G_{yy}  \mid - \sqrt{(\frac{1}{4}+G_{zz})^2-\mu^2}
    \right].
 \end{equation}
The magnetization per site $\mu$, and  the energy per site
$\varepsilon:=\frac{\la H\ra}{N}$  are obtained from the partition
function as follows:
\begin{eqnarray}\label{mu}
\mu &=&-\frac{1}{N\beta} \frac{\partial \ln Z}{\partial B},\\
\varepsilon &=& -\frac{1}{N} \frac{\partial \ln Z}{\partial \beta}
\end{eqnarray}

The correlation function $G_{zz}$ is obtained from either of the
following relations:

\begin{eqnarray}\label{corr}
G_{zz} &=&  \frac{1}{N(N-1)} \left[\frac{1}{\beta^2
Z}\frac{\partial^2 Z}{\partial  B^2}-\frac{N}{4}\right],\cr
G_{zz}&=& - \frac{1}{NJ\beta}\frac{\partial}{\partial \Delta}\ln
Z-\frac{1}{4(N-1)}.
\end{eqnarray}

The other correlation function $G_{xx}+G_{yy}$ is obtained from
\begin{equation}\label{epsilon}
    \varepsilon=\frac{\la
    H\ra}{N}=J(G_{xx}+G_{yy}+(1+\Delta)G_{zz})+\mu B,
\end{equation}
where we have used the exchange symmetry between all the pairs in
the cluster. Thus from (\ref{mu, corr}) and (\ref{epsilon}) we can
determine all the correlation functions which are necessary for the
calculation of the
concurrence. \\

The partition function is found from the following formula

\begin{equation}\label{Z}
    Z=\sum_{S=0,\frac{1}{2}}^{\frac{N}{2}} \sum_{m=-S}^S e^{-\beta \frac{J}{N-1}(S(S+1)+\Delta
    m^2)+Bm}g(N,S)
\end{equation}
where we have ignored a multiplicative constant and the lower limit
of the sum begins from $S=0(S=\frac{1}{2})$ for even (odd) $N$ and
\begin{equation}\label{gNs}
   g(N,S)=\left(\begin{array}{c} N \\
   \frac{N}{2}-S\end{array}\right) \frac{2S+1}{\frac{N}{2}+S+1}
\end{equation}
is the number of times a spin $S$ representation appears in the
decomposition of the tensor product of $N$ copies of spin
$\frac{1}{2}$ representations. Note that we have to make a
distinction between odd and even number of sites $N$, since for even
$N$ (odd $N$) only integer (half-integer) spins appear in the
decomposition. We have now set up the required equations and are
ready to analyze various situations. First we consider the isotropic
model in zero magnetic field.

\section{The xxx model in zero magnetic field}\label{xxx}  In this case
we  have $B=\Delta=0$. For a finite cluster we have no spontaneous
magnetization and hence $\mu=0$. Moreover by rotational symmetry we
have $G_{xx}=G_{yy}=G_{zz}=\frac{\varepsilon}{3J} $. Therefore we
find from (\ref{C}) that

\begin{equation}\label{Cxxx}
    C = 2 \ \ {\rm max} \ \left[ 0,\  \mid  \frac{2}{3J}\varepsilon \mid -
    \mid \frac{1}{4}+ \frac{\varepsilon}{3J}\mid \right]=2\ \  {\rm max} \ \left[ 0,\  \frac{2}{3}\mid x \mid -
    \mid \frac{1}{4}+ \frac{x}{3}\mid \right],
 \end{equation}
where we have abbreviated $\frac{\varepsilon}{J}$ to $x$. We now
note that any state with total spin $S$ has an energy given by
\begin{equation}\label{Es}
    E_S=\frac{J}{N-1}(S(S+1)-\frac{3}{4}N),
\end{equation}
from which we can derive a bound for the variable
$x=\frac{\varepsilon}{J} $. This bound is obtained by considering
the maximum and the minimum values of $S$, respectively given by
$\frac{N}{2} $ and $0 (\frac{1}{2})$ for even (odd) $N$.
\begin{eqnarray}\label{x}
-\frac{3}{4(N-1)}\leq x\leq \frac{1}{4}\ \ \ \ \ \  {\rm for} \ \ \
\ \ N&=&{\rm even}\cr -\frac{3}{4N}\leq x\leq \frac{1}{4}\ \ \ \ \ \
{\rm for} \ \ \ \ \ N&=&{\rm odd}.
\end{eqnarray}

On the other hand it is readily seen that the function
$f(x):=\frac{2}{3}\mid x \mid -
    \mid \frac{1}{4}+ \frac{x}{3}\mid
$ satisfies the following bound
\begin{equation}\label{fx}
    0< f(x)\ \ \ \ \ only\ \ if \ \ \  \  x\ < \  \frac{-1}{4}.
\end{equation}
Combination of (\ref{x}) and (\ref{fx}) shows that there is no
entanglement for any isotropic cluster for $N\geq 3$ . We can also
exclude the ferromagnetic $N=2$ cluster by noting that for such a
cluster $\varepsilon\leq 0$ (sine at infinitely high temperatures
$\varepsilon\propto tr(H) =0$, consequently at lower temperatures
$\varepsilon\leq 0$) and $J\leq 0$ making the fraction $x$ a
positive number. Thus we arrive at the result that \\

{\it{The only isotropic mean field cluster which can be
entangled at non-zero temperature is the $N=2$ anti-ferromagnetic cluster.}} \\

It had already been shown \cite{wz} that in the class of
anti-ferromagnetic rings, the $N=3$ case does not show entanglement,
here we see that this is a special case of a more general result
which holds for all mean field clusters, including $N=3$ clusters as
a special case.

\section{The xxx model in magnetic field}\label{xxxB}

\begin{figure}[t]
\centering
\includegraphics[width=8cm,height=8cm,angle=-90]{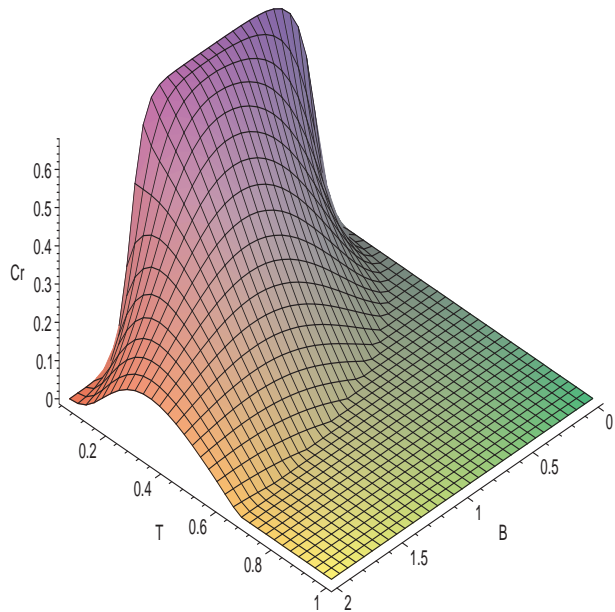}
\caption{Color online. The re-scaled concurrence for an $N=3$ xxx
cluster, a triangle of anti-ferromagnetic spins, as a function of
temperature and magnetic field. In this and all the other figures,
the units are so chosen that $B$ and $T$ become dimensionless. The
concurrence is a dimensionless quantity.}
    \label{3sites}
\end{figure}

\begin{figure}[t]
\centering
\includegraphics[width=8cm,height=8cm,angle=0]{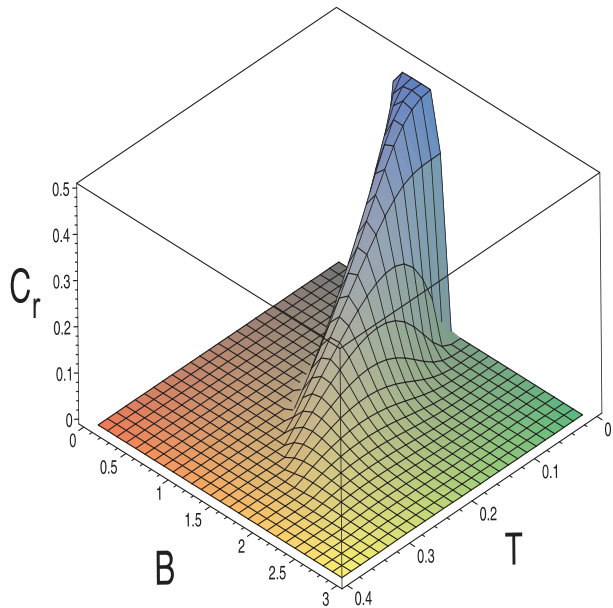}
\caption{Color online. The re-scaled concurrence for an $N=4$ xxx
cluster, a tetrahedron of anti-ferromagnetic spins, as a function of
temperature and magnetic field.}
\label{4siteslastlast}
\end{figure}

In the presence of magnetic field, the model will no longer have the
$su(2)$ symmetry and the relation $G_{xx}=G_{yy}=G_{zz}=\frac{\pm
\varepsilon}{3}$ no longer holds. However one can use (\ref{Z}) and
calculate the partition function which now takes the form

\begin{equation}\label{ZB}
   Z=\sum_{S=0,\frac{1}{2}}^{\frac{N}{2}}  e^{-\beta \frac{J}{N-1}S(S+1)}
   \frac{\sinh \beta B(S+\frac{1}{2})}{\sinh \frac{\beta
   B}{2}}g(N,S).
\end{equation}
All the required correlation functions can be obtained exactly, but
can not be written in terms of short expressions. The concurrence is
obtained numerically.  The result is that no thermal entanglement
develops in the ferromagnetic clusters but for anti-ferromagnetic
clusters, there is entanglement which generally but not always
diminishes by increasing temperature. Figures (\ref{3sites}) and
(\ref{4siteslastlast}) show the thermal entanglement of $N=3$ and
$N=4$ clusters as functions of temperature and magnetic field. In
both of them we see that there are regions in the $B-T$ plane where
an increase of temperature first increases the entanglement and then
tends to decease the entanglement. The maximum entanglement exists
at zero temperature and inside a certain interval of magnetic field
values.

\begin{figure}[t]
\centering
   \includegraphics[width=8cm,height=8cm,angle=-90]{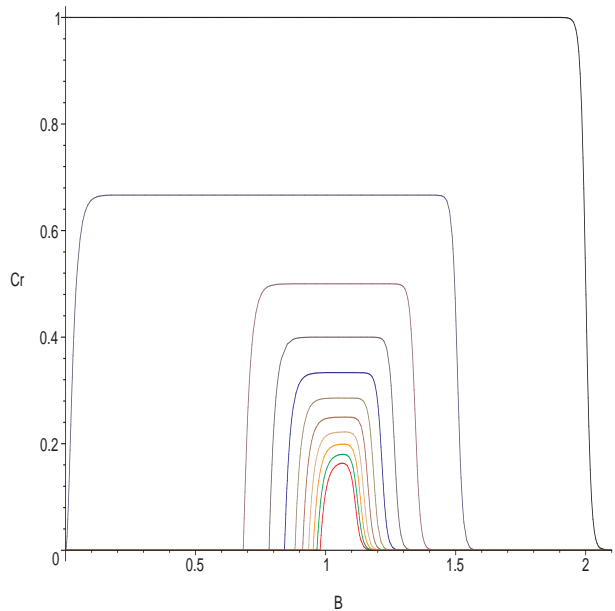}
   \caption{Color online. The re-scaled concurrence for different mean clusters at very low temperatures (T=0.01) as a function of magnetic field.
   The curves correspond (from top to bottom) to ($N=2$) to ($N=12$). The units are so chosen that $B$ and $T$ become dimensionless. The concurrence is a dimensionless quantity.}
    \label{IsotropicB}
\end{figure}

\begin{figure}[t]
\centering
   \includegraphics[width=8cm,height=8cm,angle=-90]{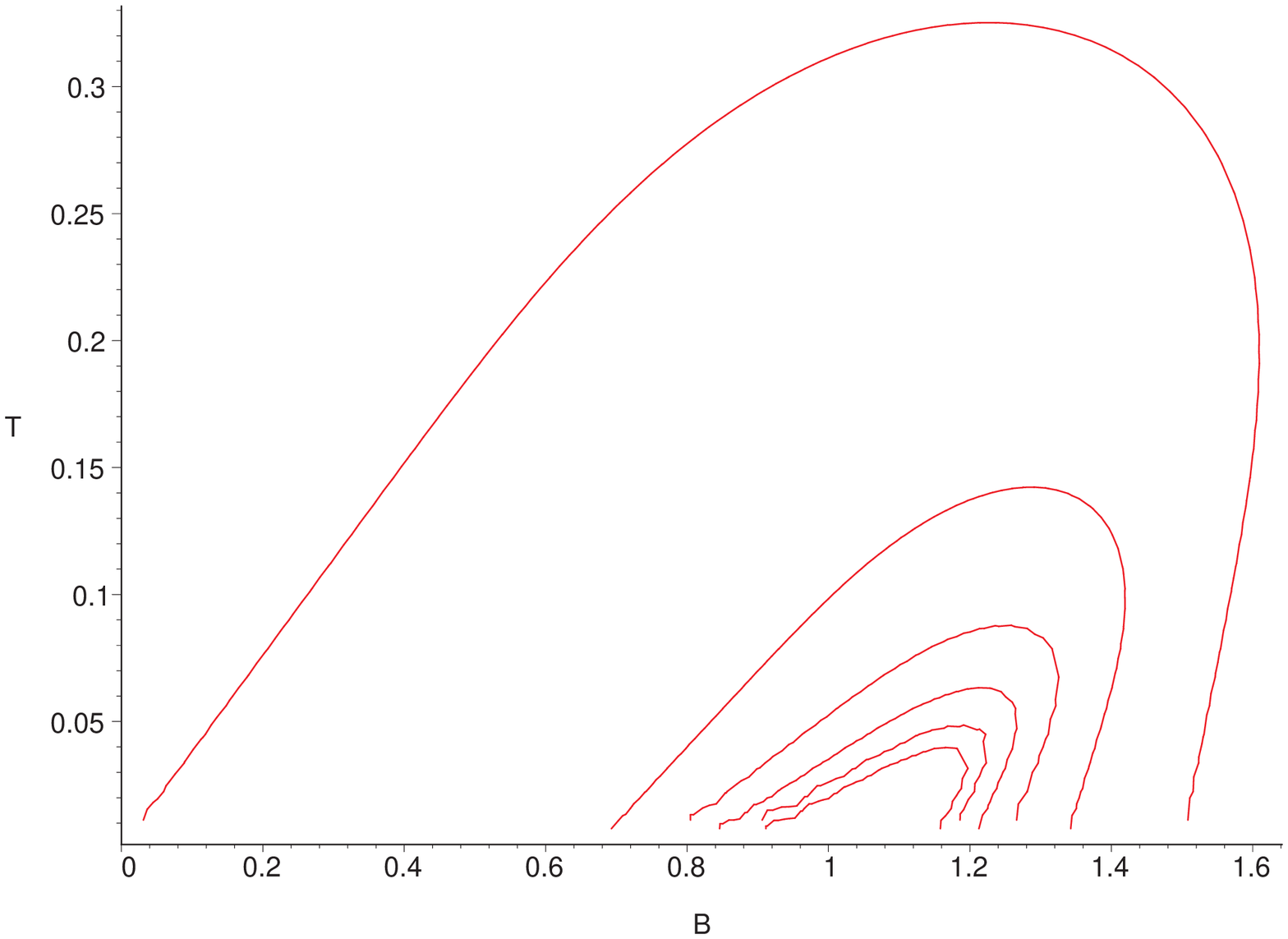}
   \caption{The regions of entanglement in the B-T plane for mean field clusters of size N=3 to size N=8
   . }
    \label{regions}
\end{figure}

In order to depict the typical information contained in these two
figures for different cluster sizes, we refer the reader to figures
(\ref{IsotropicB}) and (\ref{regions}). Figure (\ref{IsotropicB})
shows the entanglement at very low temperatures (T=0.01) for $xxx$
clusters as a function of magnetic field $B$, for cluster sizes. It
is seen that by increasing the size of clusters, the interval for
which there is entanglement shrinks to a vanishingly small interval
centered around $B=1$. Figure (\ref{regions}) shows the region in
the $B-T$ plane in which the cluster is entangled. The region
becomes smaller and smaller as the size of the cluster increases.
The right-tilted shape of these regions indicates that in certain
intervals of magnetic fields one can generate entanglement simply by
increasing the temperature. The physical reason is that when the
magnetic field is high, the ground state is a state with all the
spins aligned in the direction of magnetic field and hence there is
no entanglement at zero temperature. Slightly increasing the
temperature mixes the entangled excited states with the ground state
and generates entanglement.

\begin{figure}[t]
\centering
  \includegraphics[width=8cm,height=8cm,angle=0]{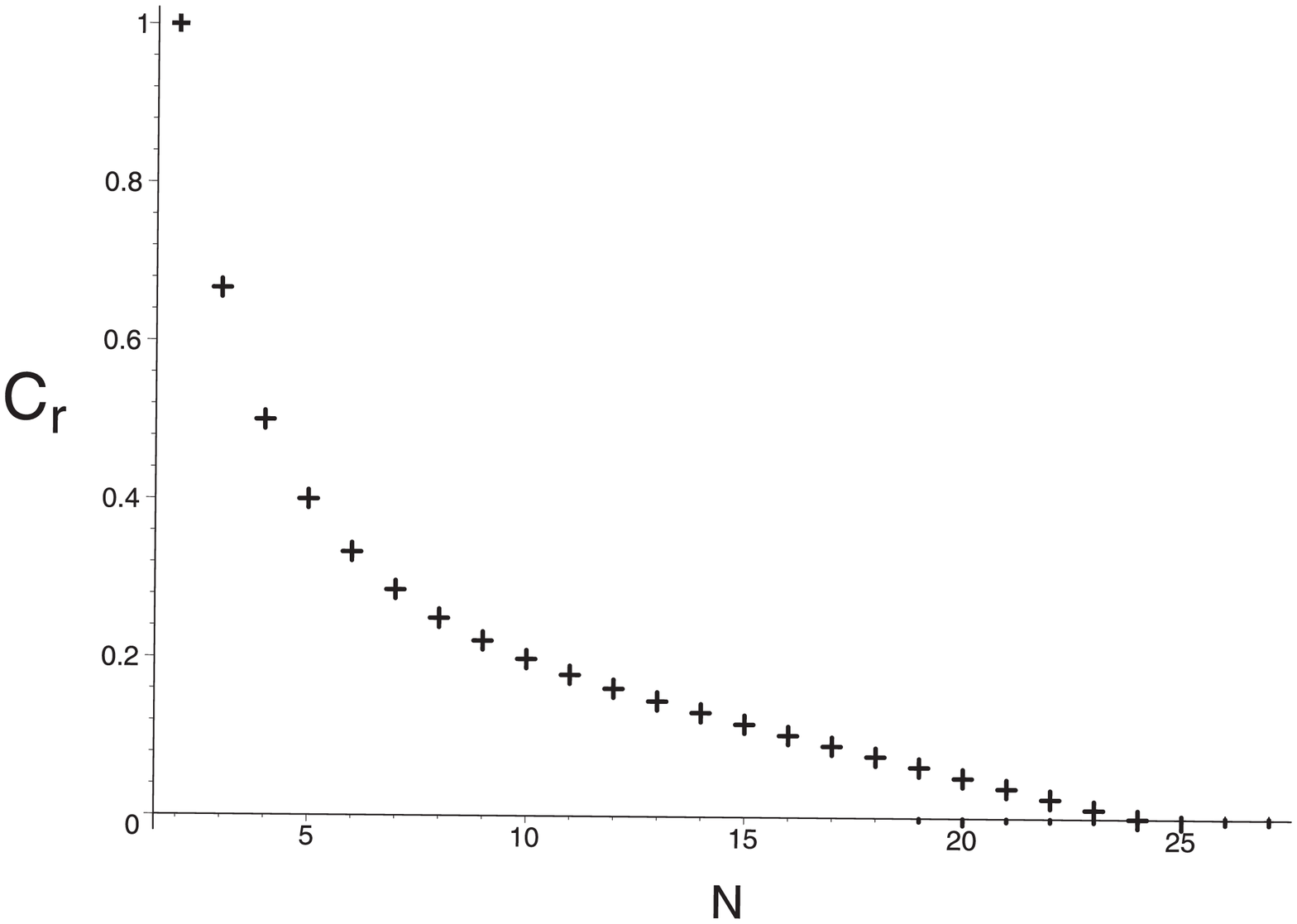}
 \caption{Maximum re-scaled concurrence for isotropic mean field clusters
in magnetic field as a function of their size $N$. For $N>23$, the
concurrence vanishes. }
    \label{pointplot}
\end{figure}

Finally figure (\ref{pointplot}) shows the maximum re-scaled
concurrence for clusters of different sizes. By re-scaled
concurrence we mean the concurrence between two sites multiplied by
$N-1$ which is the number of neighbors of a given site. It is seen
that there is almost no entanglement for clusters of size larger
than 23.

This concludes our investigation of the $xxx$ clusters in magnetic
field. We now turn to another type of anisotropy, namely the $xxz$
clusters in which  the anisotropy is not brought about by an
external magnetic field.

\section{The $xxz$ model in zero magnetic field}\label{xxz}
\begin{figure}[t]
\centering
   \includegraphics[width=8cm,height=8cm,angle=-90]{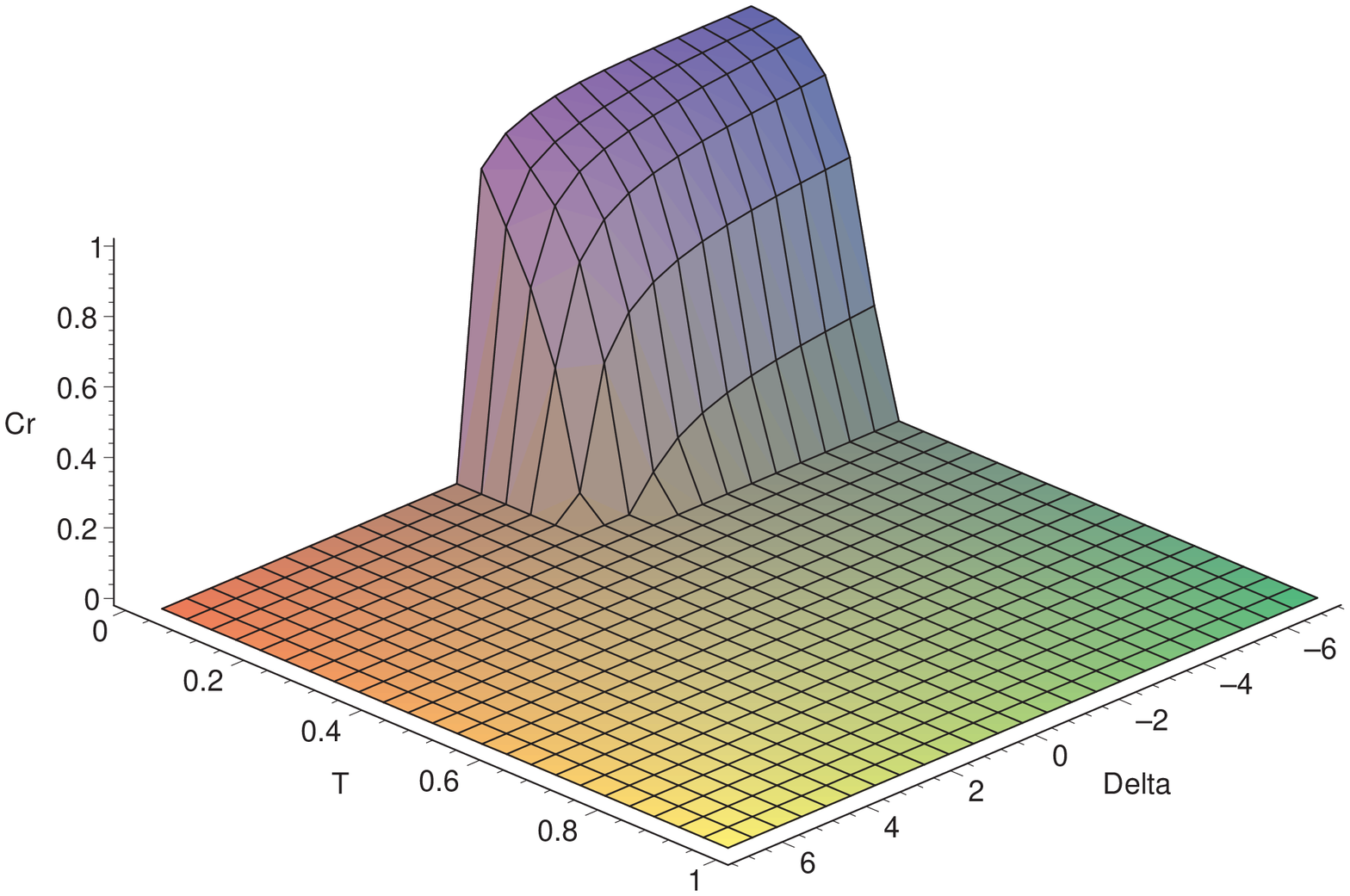}
   \caption{Color online. Re-scaled concurrence for an $N=20$ $xxz$ mean field cluster as a function of temperature and anisotropy.
    }
    \label{anisoevensample20}
\end{figure}

For these clusters we find a surprising results: Except for $N=2$,
the anti-ferromagnetic clusters show no entanglement under any
condition, regardless of the value $\Delta$ and $T$. However
ferromagnetic clusters show entanglement for all negative values of
the anisotropy parameter $\Delta$. We do not report the result for
the $N=2$ ferromagnetic cluster, since this has been reported
elsewhere \cite{rig} and focus instead on clusters of arbitrary sizes. \\
The entanglement of a ferromagnetic cluster of size $N=20$ is shown
in figure (\ref{anisoevensample20}). This figure is typical, that is
all the ferromagnetic
clusters have a similar behavior.\\

\begin{figure}[t]
\centering
   \includegraphics[width=10cm,height=10cm,angle=0]{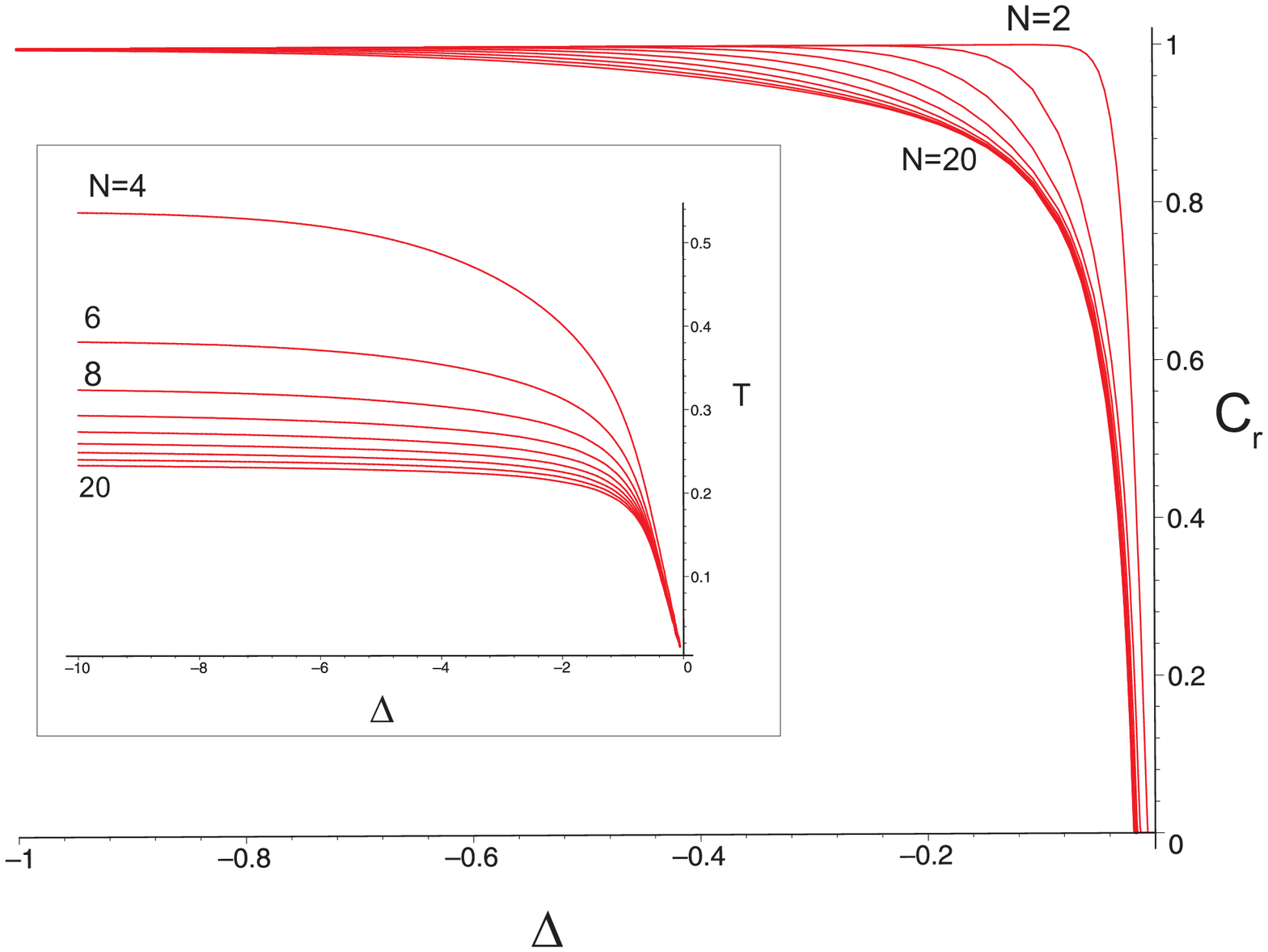}
   \caption{Re-scaled concurrence for anisotropic mean field
   clusters at nearly zero temperature (T=0.01), for clusters of even size, from N=2 to N=20. The inset shows the regions in the
   $\Delta-T$ plane below which there is entanglement. The units are so chosen that $T$ is dimensionless.
   The parameter $\Delta$ and the concurrence $C_r$ are dimensionless quantities. }
    \label{AnisoEvenCombined}
\end{figure}

\begin{figure}[t]
\centering
   \includegraphics[width=10cm,height=10cm,angle=0]{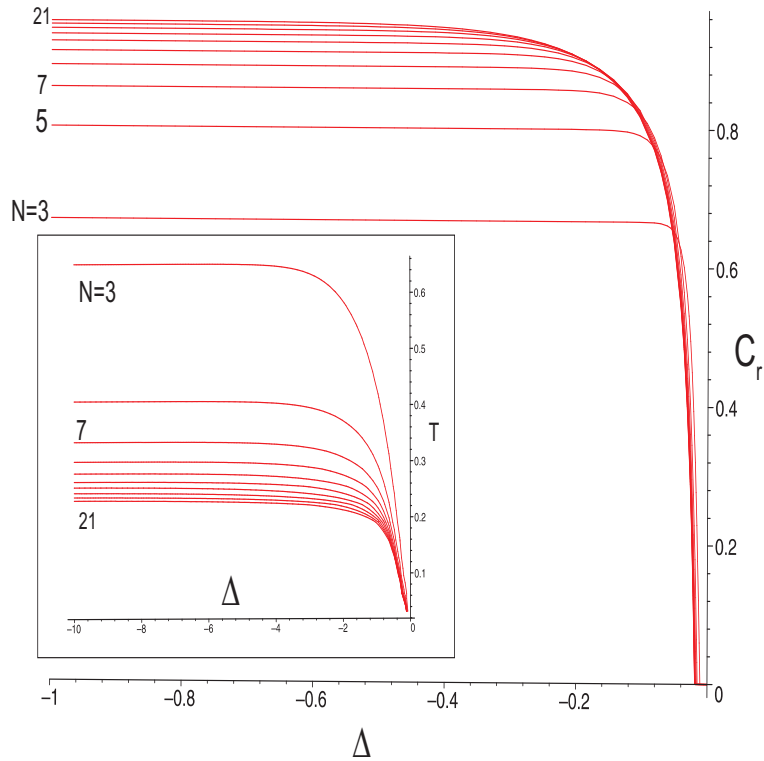}
   \caption{Re-scaled concurrence for anisotropic mean field
   clusters at nearly zero temperature (T=0.01), for clusters of odd size, from N=3 to N=21. The inset shows the regions in the
   $\Delta-T$ plane below which there is entanglement.)
   }
    \label{AnisoOddCombined}
\end{figure}

\begin{figure}[t]
\centering
   \includegraphics[width=8cm,height=8cm,angle=0]{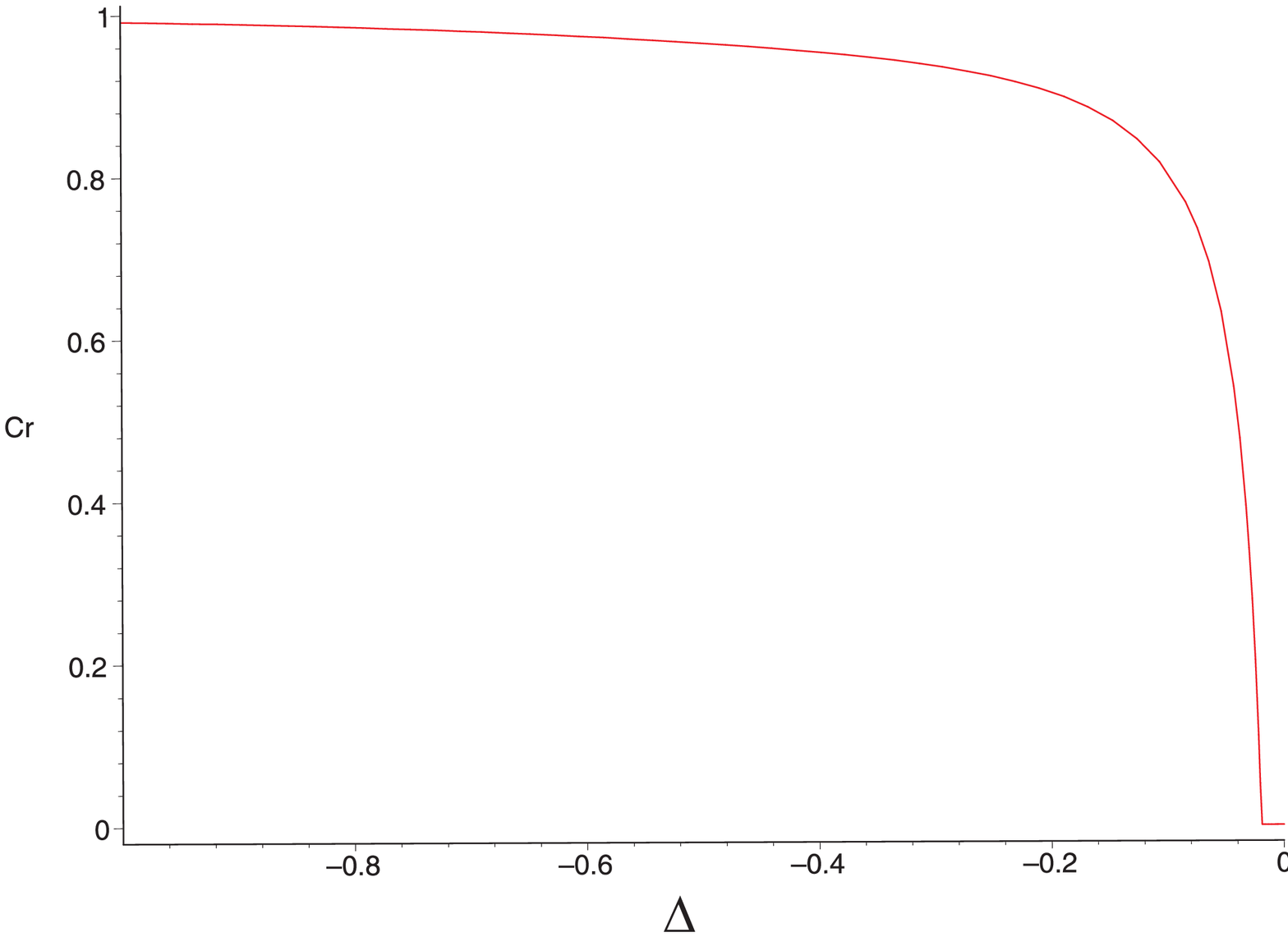}
   \caption{The re-scaled concurrence for an $xxz$ mean field cluster in the limit of
   large $N$ as a function of $\Delta$, at very low temperature $T=0.01$
   }
    \label{limit}
\end{figure}

In order to depict the relevant information for clusters of
different sizes we refer the reader to figures
(\ref{AnisoEvenCombined}) through (\ref{limit}). Figure
(\ref{AnisoEvenCombined}) and (\ref{AnisoOddCombined}) show the
entanglement at very low temperature, i.e. $T=0.01$, for even and
odd size clusters as a function of $\Delta$. Note the difference in
trends, that is, for even size clusters as $N$ increases the
re-scaled concurrence decreases while for odd size clusters, as $N$
increases the re-scaled concurrence increases. The two figures
suggest that in the limit $N\lo \infty$, the re-scaled concurrence
approaches a curve which we redraw in figure (\ref{limit}). The
insets in these figures i.e. figures (\ref{AnisoEvenCombined}) and
(\ref{AnisoOddCombined}) show the regions in the $\Delta-T$ plane
below which thermal entanglement exists for different clusters. For
both even and odd size clusters, as the size of the cluster
increases, the regions decrease in size but do not
shrink completely and approach a limiting region. \\

\begin{figure}[t]
\centering
   \includegraphics[width=10cm,height=5cm,angle=0]{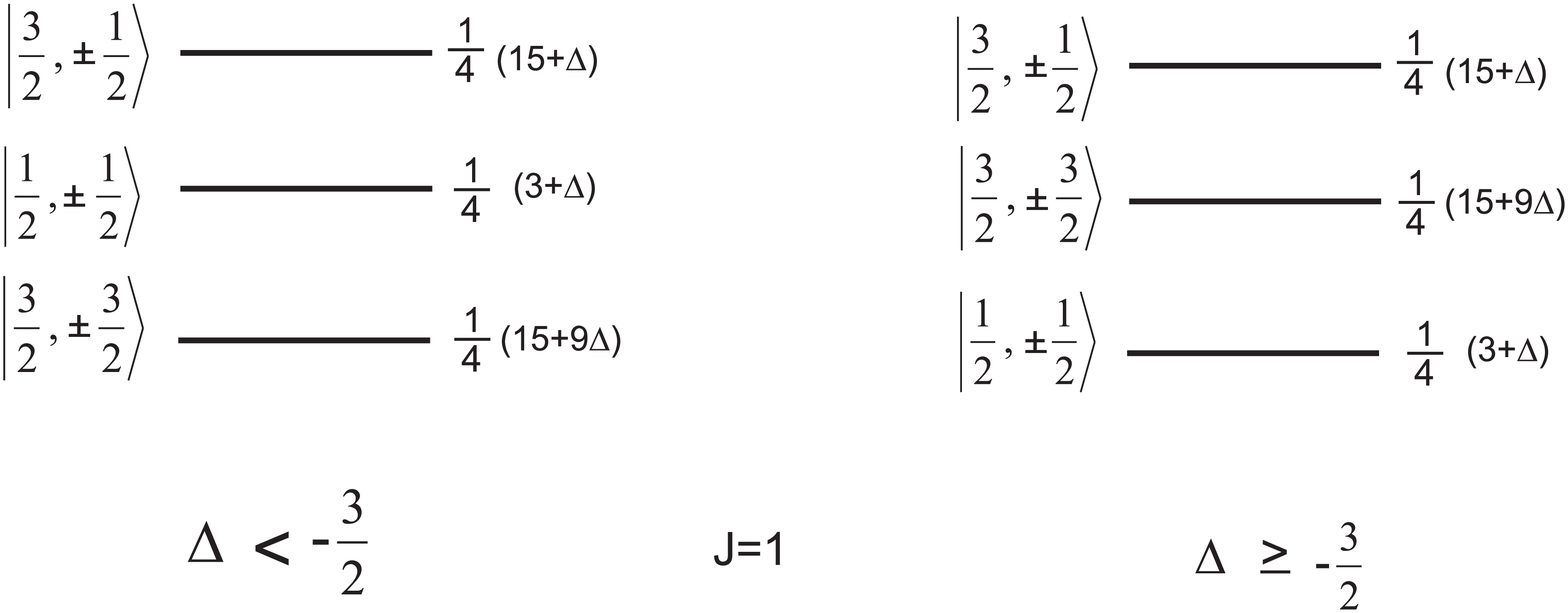}
   \caption{The spectrum of the $N=3$ $xxz$ anti-ferromagnetic
   cluster. Note that $|\frac{1}{2},\frac{1}{2}\ra$ stands for two doublets.}
    \label{antiferrospectrum_1}
\end{figure}

\begin{figure}[t]
\centering
   \includegraphics[width=10cm,height=5cm,angle=0]{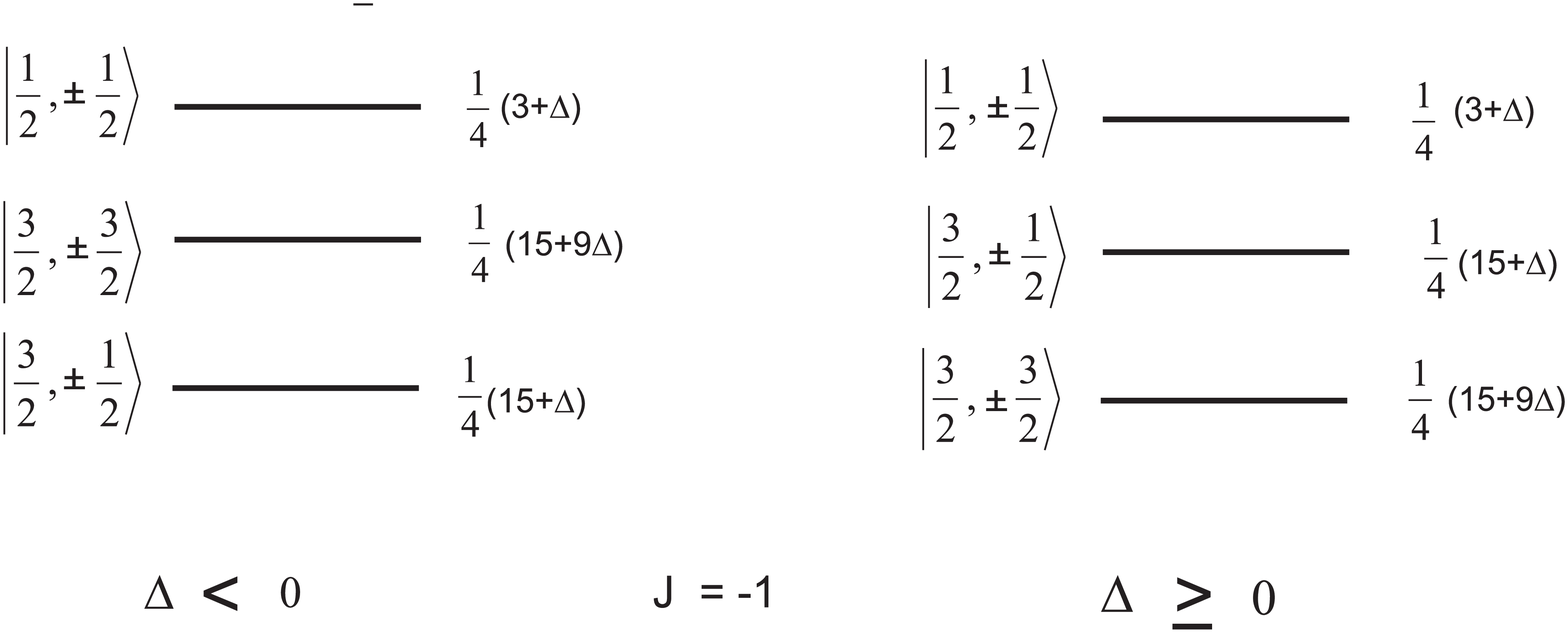}
   \caption{The spectrum of the $N=3$ $xxz$ ferromagnetic
   cluster. Note that $|\frac{1}{2},\frac{1}{2}\ra$ stands for two doublets.}
    \label{ferrospectrum_1}
\end{figure}

We have to understand two important characteristics of the
entanglement of these clusters. First why in the limit $\Delta\lo
\infty$, where the interaction approaches an Ising-like interaction
we still have entanglement at very low temperatures and second why
in the anti-ferromagnetic case there is no entanglement at all for
any value of $\Delta$. To this order we present a simple examples,
namely the $=3$ cluster.

\subsection{The N=3 cluster}

For such a cluster, the hamiltonian is
\begin{equation}\label{H3}
    H=J(s_1\cdot s_2+s_2\cdot s_3+s_1\cdot
    s_3)+J\Delta(s_{1z}s_{2z}+s_{1z}s_{3z}+s_{2z}s_{3z})\equiv
    JS\cdot S+J\Delta S_z^2,
\end{equation}
where in the second form, we have ignored an additive constant. Any
state of the form $|s,m\ra$ where $s$ and $m$ are respectively the
total and the $z$ component of the total spin is an eigenstate of
energy with energy given by $E_{s,m}=J(s(s+1)+\Delta m^2)$. The
eigenstates are as follows:

\begin{eqnarray}
  |\frac{3}{2},\ \ \frac{3}{2}\ra &=& |+,+,+\ra, \\
  |\frac{3}{2},\ \ \frac{1}{2}\ra &=& \frac{1}{\sqrt{3}}(|+,+,-\ra + |+,-,+\ra+|-,+,+\ra),\\
  |\frac{3}{2},\frac{-1}{2}\ra &=& \frac{1}{\sqrt{3}}(|-,-,+\ra + |-,+,-\ra+|+,-,-\ra), \\
  |\frac{3}{2},\frac{-3}{2}\ra &=& |-,-,-\ra,
\end{eqnarray}

\begin{eqnarray}
  |\frac{1}{2},\ \ \frac{1}{2}\ra &=& \frac{1}{\sqrt{2}}(|+,-,+\ra+|-,+,+\ra), \\
  |\frac{1}{2},\frac{-1}{2}\ra &=& \frac{1}{\sqrt{2}}(|+,-,-\ra + |-,+,-\ra),
  \end{eqnarray}

and

\begin{eqnarray}
  |\frac{1}{2}',\ \ \frac{1}{2}\ra &=& \frac{1}{\sqrt{6}}(|-,+,+\ra + |+,-,+\ra-2|+,+,-\ra),\\
  |\frac{1}{2}',\frac{-1}{2}\ra &=& \frac{1}{\sqrt{6}}(|+,-,-\ra + |-,+,-\ra+|-,-,+\ra). \\
\end{eqnarray}

The energies of the states are one of the values
$\frac{J}{4}(15+9\Delta), \frac{J}{4}(15+\Delta)$ and
$\frac{J}{4}(3+\Delta), $ depending on the values of the quantum
numbers $s$ and $m$.\\
Figures (\ref{ferrospectrum_1}) and (\ref{antiferrospectrum_1}) show
the spectrum in the ferromagnetic and anti-ferromagnetic cases
separately. In each case the nature of
spectrum depends on the value of $\Delta$. \\
Let us consider the ferromagnetic and
anti-ferromagnetic cases separately. \\

\textbf{Anti-ferromagnetic interaction: $J=1$}\\

Here we have the ground state energy $E_{gs}$ as \begin{eqnarray}
E_{gs}= \left\lbrace
  \begin{array}{l}
    \frac{1}{4}(15+9\Delta)  \ \ \ \ \ {for \ \ \ \ \ \Delta < \frac{-3}{2}},\\
\\
   \frac{1}{4}(3+\Delta)  \ \ \ \ \ \ \ {for \ \ \ \ \ \frac{-3}{2}\leq \Delta }
  \end{array}\right.
\end{eqnarray}

Consequently for $\Delta < \frac{-3}{2}$, the thermal state at zero
temperature is equal to $$\rho(T=0)= \frac{1}{2}(|+,+,+\ra\la
+,+,+|+|-,-,-\ra\la -,-,- |)$$ which is clearly a separable state.
For $\frac{-3}{2}\leq \Delta$, the thermal state is an equal mixture
of the four states with $s=\frac{1}{2}$. A simple calculation gives
in this case,  the explicit form of the two particle density matrix
as

\begin{equation}\label{rho3}
    \rho(T=0)=\frac{1}{6}\left(\begin{array}{cccc} 1 & & & \\ & 2 & -1 & \\ & -1 & 2 & \\ &&&
    1\end{array}\right),
\end{equation}

with eigenvalues $\frac{1}{6},\frac{1}{6},\frac{1}{6}$ and
$\frac{1}{2}$. Since in this case we have
$\sqrt{\rho\tilde{\rho}}:=\sqrt{\rho(\sigma_y\otimes
\sigma_y)\rho^*(\sigma_y\otimes \sigma_y)}=\rho$, we find the
concurrence of this state to be

\begin{equation}\label{C3}
    C=max\ \ (0, \frac{1}{2}-3\frac{1}{6})=0.
\end{equation}
Therefore we have shown that this anti-ferromagnetic cluster is not
entangled for any value of the anisotropy.\\
Here we have used the original form of the concurrence \cite{woo}
valid for any density matrix of two qubits which states that
\begin{equation}\label{con}
    C=max\ \ (0, \lambda_1-\lambda_2-\lambda_3-\lambda_4),
\end{equation}
where $\lambda_i$'s are the positive square roots of the eigenvalues
of $\rho\tilde{\rho}$.\\

\textbf{The ferromagnetic interaction:\ \ $J=-1$}\\

In this case the ground state energy $E_{gs}$ is
\begin{eqnarray}
E_{gs}=  \left\lbrace
  \begin{array}{l}
    \frac{-1}{4}(15+\Delta)  \ \ \ \ \ \ \ \ {for \ \ \ \ \ \Delta < 0},\\
\\
   \frac{-1}{4}(15+9\Delta)  \ \ \ \ \ \ \ {for \ \ \ \ \ 0\leq \Delta }
  \end{array}\right.
\end{eqnarray}

Consequently for $\Delta > 0$, the thermal state at zero temperature
$\rho(T=0)$  is equal to $$\rho(T=0)= \frac{1}{2}(|+,+,+\ra\la
+,+,+|+|-,-,-\ra\la -,-,- |)$$ which is clearly a separable state.
On the other hand  For $ \Delta \leq 0 $, the thermal state is an
equal mixture of the two states with $|\frac{3}{2},\frac{1}{2}\ra$
and $|\frac{3}{2},\frac{-1}{2}\ra$. A simple calculation gives in
this case,  the explicit form of the two particle density matrix as
\begin{equation}\label{rho3-a}
    \rho(T=0)=\frac{1}{6}\left(\begin{array}{cccc} 1 & & & \\ & 2 & 2 & \\ & 2 & 2 & \\ &&&
    1\end{array}\right),
\end{equation}
with eigenvalues $\frac{1}{6},\frac{1}{6},\frac{4}{6}$ and $0$. This
gives the concurrence to be

\begin{equation}\label{C3-a}
    C=max\ \ (0, \frac{4}{6}-2\frac{1}{6})=\frac{1}{3},
\end{equation}
Therefore this cluster is entangled only for negative values of
$\Delta$ and for ferromagnetic interaction. \\

Let us see why at zero or very low temperature a ferromagnetic
cluster shows entanglement in the limit of very large and negative
$\Delta$, despite the Ising like appearance of the interaction. The
reason is that in this case, the Hamiltonian approaches
$H=-(S^2+\Delta S_z^2)\approx -\Delta S_z^2$. Since $\Delta<0$, the
lowest energy states are those with $S_z=\pm \frac{1}{2}$, and these
states, as shown in figure (\ref{ferrospectrum_1}) are the states
$|\frac{3}{2},\pm\frac{1}{2}\ra $ which are highly entangled. One
may now ask why the same phenomena does not happen in the
anti-ferromagnetic case for very large positive $\Delta$, for which
the Hamiltonian takes a similar form?  The reason is that as shown
in figure (\ref{antiferrospectrum_1}) the ground state is now a {\it
mixture} of four $S_z=\pm\frac{1}{2}$ states, and this mixture does
not have any entanglement, although the states themselves may be
entangled.

\section{Discussion}
We have done a rather detailed study of the pairwise entanglement
properties of a mean field cluster of spin one-half particles
interacting via the Heisenberg $xxz$ interaction. We have considered
the ferromagnetic and anti-ferromagnetic interactions and the
isotropic ($xxx$) and anisotropic $(xxz)$ cases with and without
magnetic field. In each case we have determined the dependence of
entanglement on various parameters like the external magnetic field,
the anisotropy parameter, the cluster size as well as temperature.

It has already been shown that, although the two particle
concurrence of the ground state vanishes in the thermodynamic limit,
the re-scaled concurrence does not vanish and shows quite
interesting and nontrivial behavior \cite{vidal3, vidal4, vidal5,
vidal6}. Some of our results (figures (\ref{AnisoEvenCombined}) and
(\ref{AnisoOddCombined})) shows  that as the size of clusters
approach infinity, the re-scaled concurrence at non-zero temperature
approaches a specific function of the control parameters. In a
future work we will focus on the thermodynamic limit of these
clusters to investigate these functions where we should also take
into account the possibility of spontaneous magnetization and
symmetry breaking.
\section{Acknowledgement}
We would like to thank the members of the Quantum information group
of Sharif University for very valuable comments.

{}

\end{document}